\documentclass[preprints,article,accept,moreauthors,pdftex,10pt,a4paper]{Definitions/mdpi} 
\usepackage{lscape}

\firstpage{1} 
\pubvolume{xx}
\issuenum{1}
\articlenumber{5}
\pubyear{2018}
\copyrightyear{2018}
\history{Received: date; Accepted: date; Published: date}

\Title{Accuracy Requirements for Early Estimation of Crop Production in Senegal}

\Author{Damien C. Jacques $^{1*}$, Pierre Defourny $^{1}$}

\AuthorNames{Damien C. Jacques, and Pierre Defourny}

\address{%
$^{1}$ \quad Earth and Life Institute, Université Catholique de Louvain, 1348, Louvain-la-Neuve, Belgium
}

\corres{Correspondence: damienjacquesUCL@gmail.com}

\abstract{Early warning systems for food security rely on timely and accurate estimations of crop production. Several approaches have been developed to get early estimations of area and yield, the two components of crop production. The most common methods, based on Earth observation data, are image classification for crop area and correlation with vegetation index for crop yield. Regardless of the approach used, early estimators of cropland area, crop area or crop yield should have an accuracy providing lower production error than existing historical crop statistics. The objective of this study is to develop a methodological framework to define the accuracy requirements for early estimators of cropland area, crop area and crop yield in Senegal. These requirements are made according to (i) the inter-annual variability and the trend of historical data, (ii) the calendar of official statistics data collection, and (iii) the time at which early estimations of cropland area, crop area and crop yield can theoretically be available. This framework is applied to the seven main crops in Senegal using 20 years of crop production data. Results show that the inter-annual variability of crop yield is the main factor limiting the accuracy of pre-harvest production forecast. Estimators of cropland area can be used to improve production prediction of groundnuts, millet and rice, the three main crops in Senegal stressing the value of cropland mapping for food security. While applied to Senegal, this study could easily be reproduced in any country where reliable agricultural statistics are available.}


\begin{document}

\setcounter{section}{-1} 

\section{Introduction}

\begin{figure}[b!]
    \centering        
        \includegraphics[width=1\linewidth]{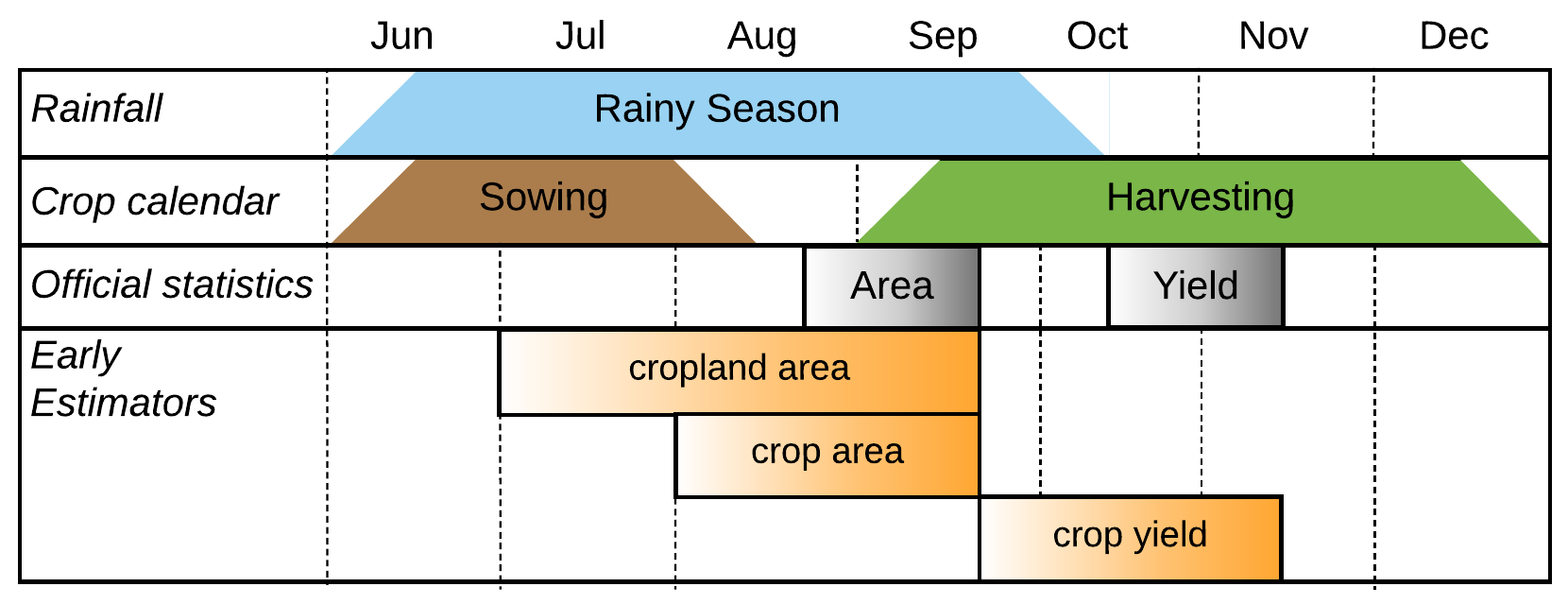} 
    \caption{Calendar of the critical periods regarding early warning systems for food security in Senegal. The rainy season lasts from June to October. Sowing starts with the first rain, and harvesting follows the end of the rainy season. Official statistics of crop area are collected from late August to the end of September, and official statistics of crop yield are collected from mid-October to mid-November. Using traditional methods such as the classification of Earth Observation data, the probability to get a good estimation of cropland area and crop area increases over the season until September when the actual value is known. On the other hand, crop yield estimations are hard to predict before the peak of the growing season in mid-September.}
    \label{fig:SchemeCalendar}
\end{figure}

Early warning systems for food security rely on timely and accurate estimations of crop production \citep{fritz2018comparison, genesio2011early, hutchinson1991uses}. Their role is to inform on risks of food crises to efficiently put in place adequate response strategies \citep{davies1991early}. Traditionally, crop statistics data are produced by national statistical offices (NSO) using adequate sampling strategies and statistical inference \citep{kreita2012global}. Final production estimates are typically available a few months after the harvest (around November in Senegal, see Figure \ref{fig:SchemeCalendar}). Yet, earlier estimation (ideally before the harvest) would substantially improve the efficiency of the political responses to food shortages. This particularly applies to countries with highly variable agricultural production such as Senegal.\\

Several approaches have been developed to get early estimations of area and yield, the two components of crop production. For crop area estimation, the most common method relies on the classification of Earth Observation (EO) data (pixel counting) following a two-step procedure: (i) an early estimate of the total cropland area and (ii) later, the discrimination of each crop within the cropland \citep{xiong2017automated,lambert2016cropland, atzberger2013advances}.  The bias of this method mainly comes from two sources of error: the presence of mixed (border) pixels and the misclassification of pure pixels \citep{Delince2017handbook}. How significant are these errors depends predominantly on the spatial resolution of the images, the spectral separability of the different crops, the classification algorithm and the quality of the training data \citep{duveiller2010conceptual, gomez2016optical, waldner2017impact, inglada2015benchmarking}. 
As for crop yield, it was shown to be correlated with vegetation indices derived from Earth Observation data \citep{tucker1985satellite, groten1993ndvi, rembold2013using, burke2017satellite, lambert2017estimate, azzari2017towards}. Agro-meteorological models such as SARRA-H \citep{sultan2005agricultural, sultan2013assessing} or GLAM \citep{challinor2004design} are also used and generally considered more rigorous. However, these models require setting several parameters for which data are often lacking (e.g. the fertilization rate).  \\


 These approaches are typically assessed regardless of the existing historical crop statistics. Yet, to be useful, early estimators of cropland area, crop area or crop yield should have an accuracy providing lower production error than existing information. For instance, having early estimators of crop area and yield providing final production estimate with an error of 25\% would not be useful if it is possible to achieve an error of only 20\% by simply using the historical average and the trend of production (information available at the beginning of the growing season). The lower the inter-annual variability of production (and the easier it is predictable by a trend), the higher would be the accuracy requirements for early estimators of cropland area, crop area and crop yield. In the Sudano-Sahelian zone, this variability is generally significant due to the high inter-annual change in rainfall pattern specific to the region \citep{graef2001spatial, grist2001study, haarsma2005sahel}. Several factors can also play a role such as agronomic conditions (soil quality), crop management practices (fertilization rate, phytosanitary protection), pest outbreaks (locust, birds), infrastructure development (storage facilities) as well as market conditions (price, demand, competition) \citep{Sghir2015, d2015senegal}. \\

From an early warning perspective, a critical factor that should be taken into account is the date on which ground truth data are available. In Senegal, official statistics of crop areas are known at the end of September and crop yield at the end of November (Figure \ref{fig:SchemeCalendar} - Official statistics). It is unlikely to get accurate estimations of yield earlier than September as the end of the vegetation growth, a critical period in determining the final yield, occurs at the same time (Figure \ref{fig:MapProd} - NDVI zones). Indeed, most of the study estimating crop yield from EO data either use the maximum of a vegetation index such as the Normalized Difference Vegetation Index (NDVI) \citep{groten1993ndvi,lambert2017estimate, azzari2017towards, becker2010generalized} or its integration over the growing cycle \citep{tucker1985satellite, boken2002improving}. Practically, it means that early estimators of cropland and crop area are useful before September and estimators of yield, between September and November  (Figure \ref{fig:SchemeCalendar} - Early Estimators). \\

The overall objective of this study is to develop a methodological framework defining the accuracy requirements for early estimators of cropland area, crop area and crop yield in Senegal. These requirements are made according to (i) the inter-annual variability and the trend of historical data, (ii) the calendar of official statistics data collection, and (iii) the time at which early estimations of cropland area, crop area and crop yield can theoretically be available (based on EO data). This framework is applied to the seven main crops in Senegal using 20 years of crop production data.  While applied to Senegal, this study could easily be reproduced in any country where reliable agricultural statistics are available.

\section{Agriculture in Senegal}

\begin{landscape}
\begin{figure}
    \centering        
        \includegraphics[]{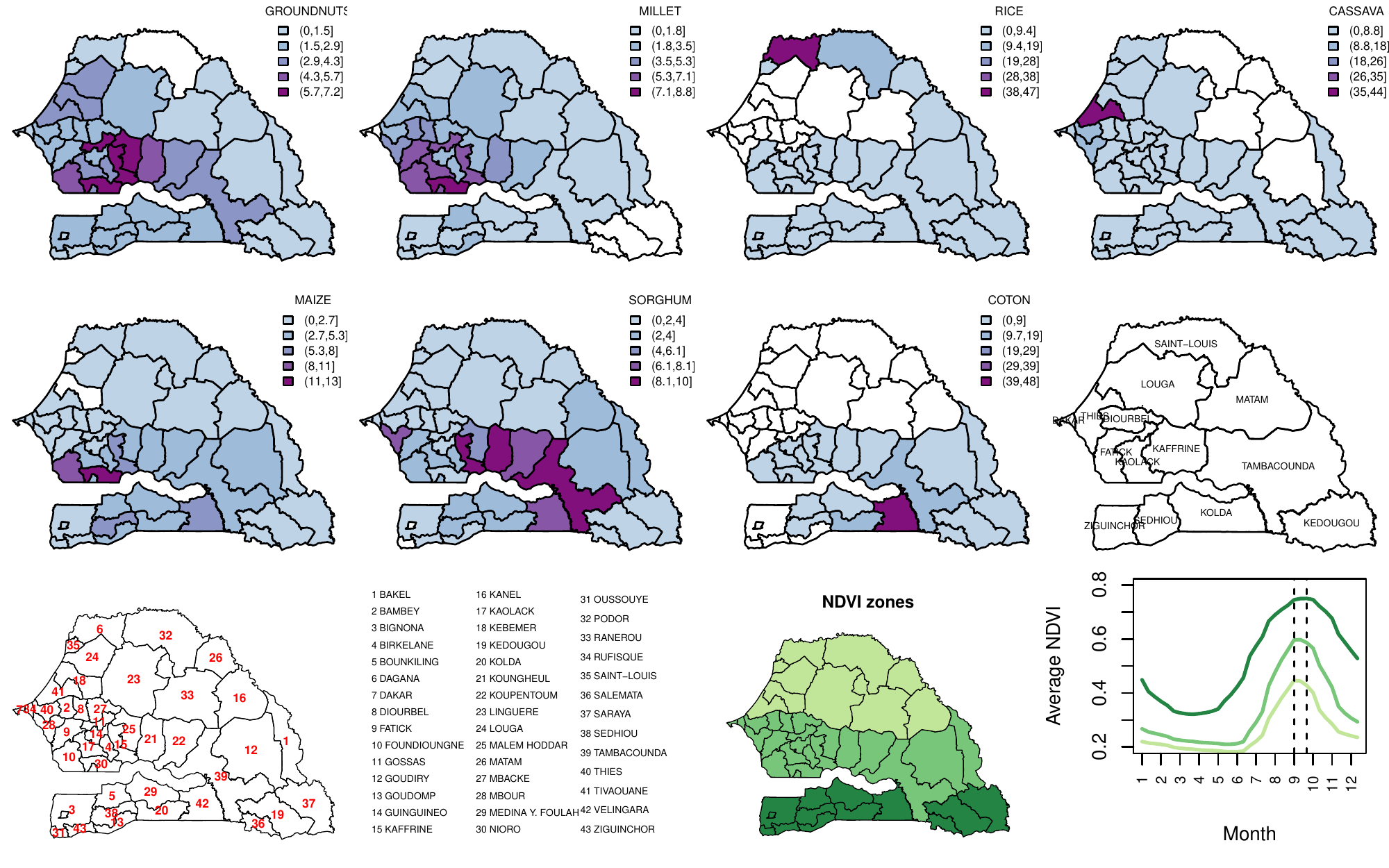} 
    \caption{The two top rows show the average production (2010-2016, source: DAPSA) by department for each crop in percentage of the total production (white areas correspond to null production) and a map of the 14 regions in Senegal. The last row shows a map of the department with its legend and the average NDVI (from MODIS images) in 3 zones following a South-North gradient. The dashed lines on the last plot show the period of maximum NDVI (end of September and beginning of October). }
    \label{fig:MapProd}
\end{figure}
\end{landscape}

Today, the agriculture sector accounts for one-sixth of the Senegalese gross domestic product \citep{d2015senegal}. Rice is the main staple food consumed by Senegal’s population, but production is only able to meet half of the demand \citep{CGER2014}. Because of its importance in Senegalese food, rice self-sufficiency has been the target of public authorities for many years. Following the 2007-2008 food crisis, several public initiatives such as the \textit{Programme National d’Autosuffisance en Riz} (PNAR) or the \textit{Grande Offensive Agricole pour la Nourriture et l’Abondance} (GOANA), helped to increase the production over the years. Most of the rice is irrigated and produced in Senegal river valley by the \textit{Société Nationale d’Aménagement et d’Exploitation des Terres du Delta du fleuve Sénégal et des vallées du fleuve Sénégal et de la Falémé} (SAED) and the \textit{Société de Développement agricole et industriel} (SODAGRI) (Figure \ref{fig:MapProd}). \\

\begin{figure}
    \centering        
        \includegraphics[width=0.75\textwidth]{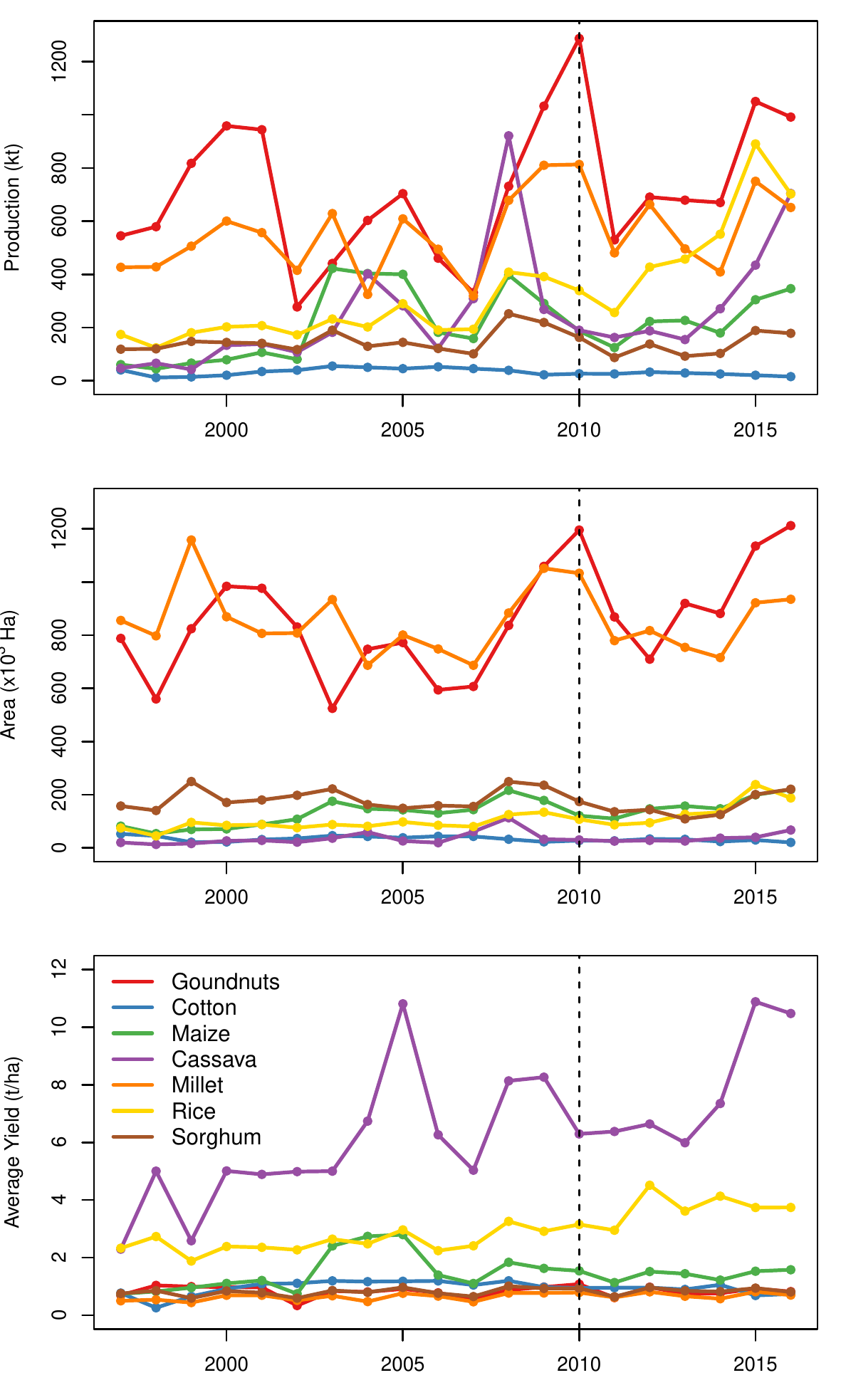} 
    \caption{National production, area and yield time-series of each crop. The vertical dotted line shows the date upon which departments level data were available.}
    \label{fig:LTProd}
\end{figure}

Millet, and to a lesser extent, sorghum, serves as the main local subsistence food crops~\citep{dong2011}. Millet is the most drought-resistant crop in the country, and it covers a third of Senegal’s cultivated areas (Figure~\ref{fig:LTProd}). Most of the millet is grown in the regions of Kaolack, Kaffrine and Fatick where it is interchanged with groundnuts (Figure~\ref{fig:MapProd}). This crop rotation is crucial because groundnuts fix nitrogen in the soil. \\

\begin{figure}
    \centering        
        \includegraphics[width=0.65\linewidth]{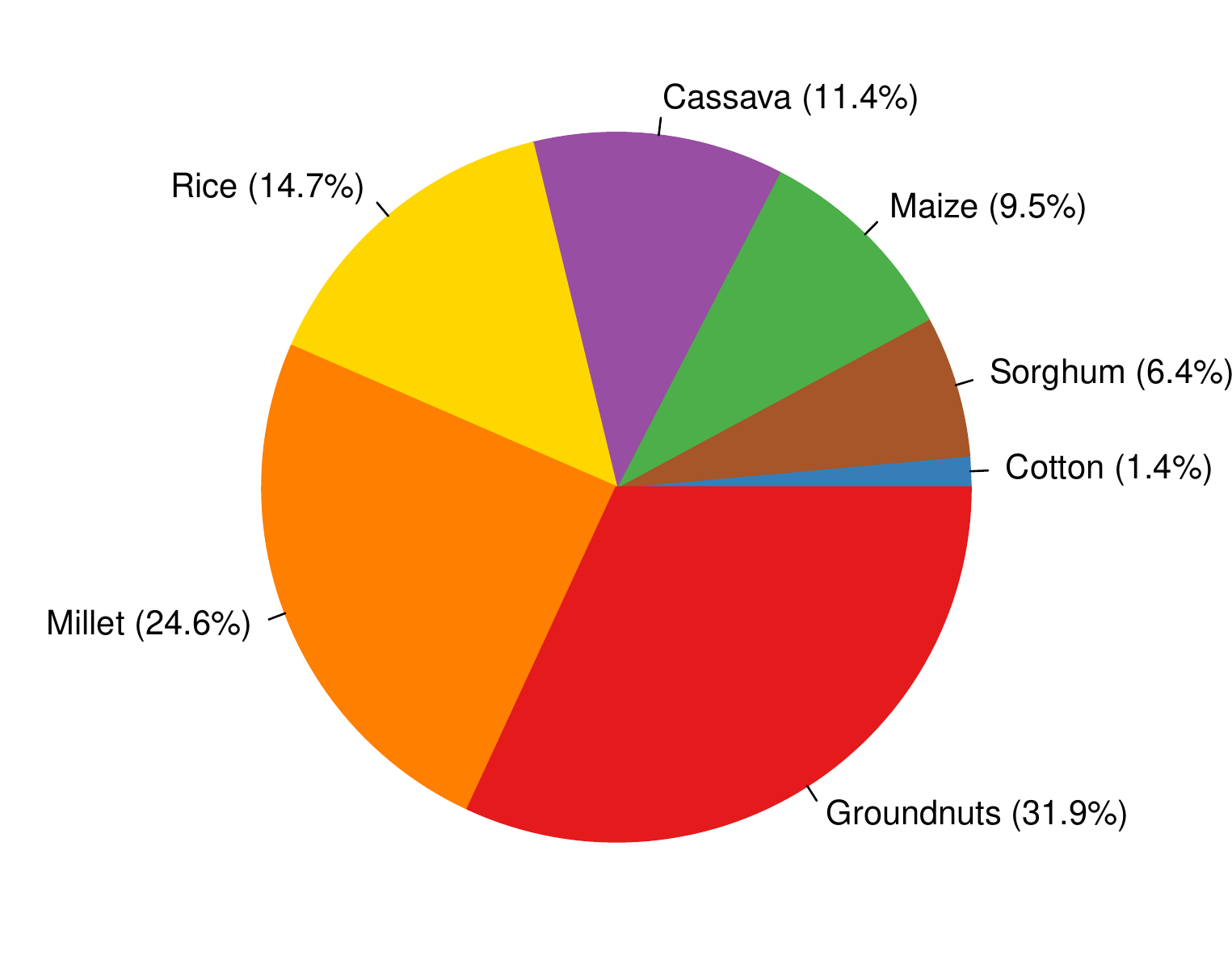} 
    \caption{Pie chart of the average national production (1997-2016) of the seven main crops in Senegal.}
    \label{fig:Pie}
\end{figure}

Groundnuts are the main cash crop of the country. While the sector is going through a crisis since the nineties due to several factors such as the different agricultural policies, the market fluctuation or the deterioration of farm equipment, it is still the first production in Senegal (Figure \ref{fig:Pie}) and it plays a vital role in rural economy \citep{noba2014arachide}. Groundnuts are primarily cultivated in the Groundnuts Basin located in the region of Fatick, Kaolack, and Kaffrine.\\

Cotton is the second cash crop in the country (Figure~\ref{fig:Pie}). The production is low and almost entirely exported \citep{isracirad2005}. In Senegal, most of the cotton is produced by the \textit{Société de Développement et des Fibres Textiles} (SODEFITEX) in the region of Kolda.\\

Cassava and maize have raised public policies interest as an interesting source of diversification for the subsistence agriculture in Senegal. In 2004, showing an aggressive agricultural policy and revived interest, the Senegalese government launched a major program, the \textit{Programme Spécial de Relance de la Filière Manioc au Sénégal} (PSRFMS), for intensifying the production of cassava to improve food security \citep{diallo2013importance}. The cassava and maize sector have also benefited from the GOANA support in 2008. Cassava is mainly cultivated in Thiès region, and maize, in Fatick and Kaolack region (Figure \ref{fig:MapProd}). \\

Other main crops in Senegal include sesame (12 kt in 2016), watermelon (285 kt in 2016) and cowpea (100 kt in 2016). These are not taken into account in this study.

\section{Data}

Official agricultural statistics delivered by the \textit{Direction de l'Analyse, de la Prévision et des Statistiques Agricoles} (DAPSA) are used in this analysis \citep{DAPSA2017}. In this study, these are regarded as the ground truth and serve as the reference to evaluate the inter-annual variability of production. It is not a strong assumption as the statistical capacities are rather good in Senegal compared to the African average. According to the statistical capacity score\footnote{The Statistical Capacity Indicator is a composite score assessing the capacity of a country’s statistical system. It is based on a diagnostic framework evaluating the following areas: methodology; data sources; and periodicity and timeliness. Countries are scored against 25 criteria in these areas, using publicly available information and/or country input. The overall Statistical Capacity score is then calculated as a simple average of all three area scores on a scale of 0-100.} computed by the World Bank, Senegal is classified second among all sub-Saharan Africa countries in 2017 (\url{datatopics.worldbank.org/statisticalcapacity}). \\

\begin{table}
\centering
\begin{tabular}{lrrrr}

&  \multicolumn{4}{c}{Production (kt or \%)}  \\ 
  & min & mean & max & CV  \\ 
    \midrule
 Groundnuts & 277.3 & 715.9 & 1286.9 & 36.8  \\ 
  Millet & 318.8 & 552.3 & 813.3 & 26.8 \\ 
   Rice & 123.5 & 329.4 & 890.4 & 60.5  \\ 
 Cassava & 42.1 & 255.6 & 920.9 & 86.2  \\ 
 Maize & 44.3 & 213.8 & 422.0 & 60.2  \\ 
 Sorghum & 86.9 & 144.5 & 251.5 & 29.7  \\ 
  Cotton & 11.6 & 32.0 & 55.0 & 41.3  \\ 
    \bottomrule 
    \\
    &   \multicolumn{4}{c}{Area ($10^{3}$ ha or \%)} \\ 
  & min & mean & max & CV \\ 
    \midrule
 Groundnuts &  524.8 & 850.8 & 1211.0 & 23.7  \\ 
  Millet &  686.9 & 852.3 & 1158.2 & 14.6 \\ 
   Rice &  45.2 & 106.6 & 238.2 & 40.6  \\ 
 Cassava & 13.1 & 36.3 & 113.2 & 64.2  \\ 
 Maize &  53.7 & 135.2 & 218.5 & 35.5  \\ 
 Sorghum &  108.8 & 177.0 & 249.7 & 23.4  \\ 
  Cotton &  20.6 & 33.6 & 52.6 & 28.3  \\ 
    \bottomrule 
    \\
    &   \multicolumn{4}{c}{Yield (t/ha or \%)}\\ 
  & min & mean & max & CV  \\ 
    \midrule
 Groundnuts &  0.33 & 0.83 & 1.08 & 21.7 \\ 
  Millet &  0.44 & 0.64 & 0.81 & 18.8 \\ 
   Rice &  1.88 & 2.93 & 4.51 & 24.2 \\ 
 Cassava & 2.29 & 6.46 & 10.97 & 37.0 \\ 
 Maize &  0.74 & 1.47 & 2.80 & 40.1 \\ 
 Sorghum &  0.59 & 0.82 & 1.01 & 15.9 \\ 
  Cotton &  0.26 & 0.95 & 1.19 & 25.3 \\ 
    \bottomrule 
\end{tabular}
  \caption{Descriptive statistics of national production, area, and yield for each crop over 20 years (1997-2016). CV stands for Coefficient of Variation computed as the standard deviation divided by the mean (expressed in \%) which is a measure of the inter-annual variability (without taking into account for the trend).}
  \label{tab:statDesc}
\end{table}

The estimates are based on a two-stage stratified sample of 6300 agricultural households in accordance with an harmonized approach intended to be applied similarly in all countries of the \textit{Comité inter-États de lutte contre la sécheresse au Sahel} (CILSS). The sample size has slightly changed over the years, but the method remains the same. As a first stage, between 18 and 30 census districts (CDs) are randomly drawn in each of the 42 departments of Senegal (Dakar is not included) in proportion to their population for a total of 900 sampled CDs out of 17,165. Then, within each CD, 7 agricultural households are randomly drawn which giving a final sample of 126 to 210 households by department. \\

The data collection is divided into two phases (Figure \ref{fig:SchemeCalendar}). During Phase 1, which starts at the end of August, the crop area is estimated. The crop type and the location of all fields of each sampled household are recorded. Phase 2 begins in October at the end of the growing season. At that time, the yield is estimated by averaging measurements taken within 60 plots for each crop and department in a sample of the monitored fields (dimension specific for each crop, for instance, 5m x 5m for groundnuts and 1m x 1m for rice). Both phases last 30 days meaning that the crop areas are known in September and the crop yields, in November. Some crops are not monitored by the DAPSA and their statistics are provided by other institutions. These are the rainy season irrigated rice in SAED (The regions of St-Louis and Matam and the department of Bakel) and SODAGRI (Anambé and Velingara department) zone, and the cotton in SODEFITEX zone. \\

The seven main crops of Senegal are considered in this study: groundnuts, millet, rice, cassava, maize, sorghum and cotton (Figure~\ref{fig:Pie}, Table~\ref{tab:statDesc}). Rice referred to paddy rice, when husked, it loses about 30\% of its weight. National production data were available from 1997 to 2016 and local production (at department level) were available from 2010 to 2016. Both data were obtained directly from the DAPSA. \\


\section{Method}


The objective of this paper can be further defined in specific research questions listed hereafter.\\

First, based on twenty years of national agricultural production: 

\begin{enumerate}
\item What part of the inter-annual variability of crop production is predictable by the trend?
\end{enumerate}

Second, based on seven years of departmental agricultural production: 

\begin{enumerate}
  \setcounter{enumi}{1}
\item How the inter-annual variability of crop production is spatially distributed? 
\end{enumerate}

Finally, according to (i) the inter-annual variability of production and the trend, of historical data (first research question) (ii) the calendar of official statistics data collection (Figure \ref{fig:SchemeCalendar} - Official Statistics), and (iii) the time at which early estimations of cropland area, crop area and crop yield can theoretically be available based on EO data (Figure \ref{fig:SchemeCalendar} - Early Estimators):

\begin{enumerate}
  \setcounter{enumi}{2}
\item  What is the lowest error of crop production estimation achievable along the season?
\item What are the accuracy requirements of early estimators of cropland area, crop area and crop yield?
\end{enumerate}


The following sections detail the methodological background used to answer these research questions.

\subsection{Production Estimator and Error Measurement}

For a crop $i$ and a year $t$, the error between the actual production $p_{it}$ and the production estimation $\hat{p}_{it}$ is given by:

\begin{equation}
\epsilon_{it} = \hat{p}_{it} - p_{it}
\label{Eq:error}
\end{equation}

A good estimator should provide low error $|\epsilon_{it}|$ with $\epsilon_{it} = 0$ for a perfect estimator. An estimator can give good estimates for a year $t$ but poorly estimates the year $t + 1$. For this reason, the quality of an estimator should ideally be assessed over several years. From a vector of predictions $\epsilon_{it}$, ...,$\epsilon_{iT}$ (where $T$ is the number of years in the time series), several criteria can be chosen as indicator of quality, the most simple being the average error $\bar \epsilon_{i}$ of each year $t$. One could also define a threshold value that $|\epsilon_{it}|$ should never exceed. For instance, an estimator would be considered insufficient if one year is estimated with an error exceeding 30\%, even if the average error of all years $\bar \epsilon$ is below this threshold. In this study, we opted for the Root Mean Square Error (RMSE) as the indicator of quality of a crop production estimator:

\begin{equation}
\mathrm{RMSE}_{i}=\sqrt{\frac{\sum_{t=1}^T \epsilon^2}{T}}
\label{Eq:RMSE}
\end{equation}

Compared to a simple average $\bar \epsilon_{i}$, RMSE penalizes large errors. To be able to compare RMSE of different crop, $\mathrm{RMSE}_{i}$ was normalized by the average production $\bar{p}_{i}$ which gives the coefficient of variation $\mathrm{CV(RMSE_{\textit{i}})}$:

\begin{equation}
\mathrm{CV(RMSE_{\textit{i}})} = \frac {\mathrm{RMSE}_{i}}{\bar{p}_{i}} 
\label{Eq:CVRMSE}
\end{equation}

A good estimator will have a low CV(RMSE) while a poor estimator will have a large CV(RMSE).\\

\subsection{Average of Past Data}

Without an apparent trend, the production (at country level) of a crop $i$ and a year $t$ can be estimated by the average of past production data:

\begin{equation}
\hat{p}_{may, it|T} = \bar{p}_{i} = (p_{i1}+...+p_{iT})/T
\label{Eq:prodAv}
\end{equation}

\noindent where $p_{i1},...,p_{iT}$ are the past production of crop $i$. Because it only uses past data, this estimator is available at the beginning of the growing season (in May in Senegal). \\

To calculate the CV(RMSE) of this estimator, we used historical data spanning twenty years from 1997 to 2016. We computed the error of estimation for each year $t$ using the average of all other years $T$ (19 years) as predictor. This approach corresponds to a leave-one-out cross validation (LOOCV). The CV(RMSE) will be large if the inter-annual variability of production (the CV of the time series reported in Table \ref{tab:statDesc}) of the crop is high. Conversely, if the inter-annual variability of the crop production is low, the CV(RMSE) will be low and the accuracy requirements of early estimators of cropland, crop area and crop yield will be high. \\ 

Using an approach similar to Eq.~\ref{Eq:prodAv}, at the beginning of the season, we can estimate the area  ${a}_{it}$ and the yield ${y}_{it}$ using $\hat{a}_{it|T}$ and $\hat{y}_{it|T}$ respectively. Furthermore, an estimator of the proportion of crop area in total cropland area $\widehat{(a/c)}_{it|T}$ can also be computed using historical average of $a_{it}/c_{t}$ where $c_{t}$ is the total cropland area of year $t$. This last estimator can be interesting when the total cropland area of the predicted year is the only information available (Figure \ref{fig:SchemeCalendar} -- Early Estimators).  \\

These four estimators $\hat{p}_{it|T}$, $\hat{a}_{it|T}$, $\hat{y}_{it|T}$ and $\widehat{(a/c)}_{it|T}$ are available at the beginning of a new growing season (in May in Senegal) and form the baseline for the accuracy requirements of early estimators of cropland area, crop area and crop yield computed from, for instance, EO data.

\subsection{Trend of Past Data}

In temperate climate and developed countries, most of the production variability can be explained by the trend because long-term policies, market change as well as technology and practices improvements are the driving force of production growth \citep{defourny2007respective}. In the Sudano-Sahelian zone, it is not the case as unpredictable events such as droughts, floods, pest outbreaks (locust, birds), limited access to inputs or conflicts have a large effect on production variability and long-term policies are not always efficient \citep{ d2015senegal}. Yet, supporting programs and economic investments in a specific sector may still result in a sustained increase in production. To identify such a trend, simple linear (Eq.~\ref{Eq:DetrendLinear}) and exponential (Eq.~\ref{Eq:DetrendExpo}) regression models (ordinary least square) were run with national production, area, and yield against time for each crop: 

\begin{equation}
{u}_{it}=\alpha_{i}+\beta_{i}{t}+{\epsilon}_{it}
\label{Eq:DetrendLinear}
\end{equation}

\begin{equation}
log({u}_{it})=\alpha_{i}+\beta_{i}{t}+{\epsilon}_{it}
\label{Eq:DetrendExpo}
\end{equation}

\noindent where ${u}_{it}$ is the production ${p}_{it}$, the area ${a}_{it}$ or the yield ${y}_{it}$ of crop $i$ and year $t$, and $\epsilon$ is the regression residual. We assumed that there was an actual trend in a time series when the $\beta_{i}$ coefficient was statistically significant ($p$-value$<$0.01). In the case where both models (linear and the exponential) had a significant $\beta_{i}$, the model with the highest coefficient of determination R$^{2}$ was selected.\\ 

As theoretically predictable and, therefore, known at the beginning of the growing season, significant trends were subtracted from the historical time series:

\begin{equation}
{z}_{it}={u}_{it} - \text{trend}_{it} 
            \quad \text{with} \quad \text{trend}_{it}= \begin{cases}  
             \alpha_{i}+\beta_{i}{t} \quad \text{if the trend is linear}, \\ 
              e^{\alpha_{i}+\beta_{i}{t}} \  \quad \text{if the trend is exponential}\\  
       \end{cases} 
\label{Eq:Detrend}
\end{equation}

\noindent where ${z}_{it}$ is the detrended production, the detrended area or the detrended yield of crop $i$ and year $t$. \\

For the crops with a significant trend, Eq.~\ref{Eq:prodAv} can then be rewritten as:

\begin{equation}
\hat{u}_{may, it|T} = \bar{z}_{i} + \text{trend}_{it} 
\label{Eq:prodEq}
\end{equation}

\noindent where $\bar{z}_{i} = (z_{i1}+...+z_{iT})/T$ with $z_{i1},...,z_{iT}$ the detrended past production, area  or yield  of crop $i$ as defined in Eq.~\ref{Eq:Detrend}.\\

\subsection{Spatial Variability}

Production variability is a function of time but also space. Some departments have a stronger variability than others. Only seven years of historical data were available at the department level (2010-2016) which was not enough for a rigorous analysis. However, a general picture can be derived from $\hat{p}_{may, it|T}$ accuracy for each department and each crop. For this analysis, the trend was not taken into account as hard to accurately estimate on short time series. The spatial analysis of production variability has interesting operational implication because it allows to broadly stratifying the country in priority zones. Departments that contribute the most to the national variability should be targeted (sampled) first to get accurate estimations of production.

\subsection{Early Estimators}

Along the season, early estimators of cropland area $\hat{c}_{it}$, crop area $\hat{a}_{it}$ and crop yield $\hat{y}_{it}$ become available thanks to EO data (see Figure \ref{fig:SchemeCalendar} -- Early Estimators). \\

In July/August, crops start growing. Using EO data, it is still very difficult to distinguish between each crop but the entire area covered by the cropland can be estimated. The production is estimated as:

\begin{equation}
\hat{p}_{jul,it} = \hat c_{t}\widehat{(a/c)}_{it|T}\hat{y}_{it|T}
\label{Eq:prodJul}
\end{equation}

In August/September, crops have now sufficiently grown to be identified. Using EO data, the area of each crop can be estimated. However, it is still too early to predict the crop yield. The production is estimated as:

\begin{equation}
\hat{p}_{aug,it} = \hat{a}_{it}\hat{y}_{it|T}
\label{Eq:prodAug}
\end{equation}

In September,  the growing season has reached its peak and the official statistics of crop area ${a}_{it}$ are known. The production is estimated as:

\begin{equation}
\hat{p}_{sep,it} = {a}_{it}\hat{y}_{it|T}
\label{Eq:prodSep1}
\end{equation}

After September,  the growing season is over and the crop yield can be estimated using EO data. The production is estimated as:

\begin{equation}
\hat{p}_{oct,it} = {a}_{it}\hat{y}_{it}
\label{Eq:prodSep2}
\end{equation}

Finally in November, official statistics of crop yield ${y}_{it}$ are known and the final estimate of production (considered the ground truth) is available:

\begin{equation}
{p}_{nov, it} = {a}_{it}{y}_{it} 
\label{Eq:prodNov}
\end{equation}

The time at which ${p}_{may}$, ${p}_{sep}$ and ${p}_{nov}$ are available is defined by the data collection schedule of the crop statistics office (Figure \ref{fig:SchemeCalendar} -- Official Statistics). On the other hand, the dates for ${p}_{jul}$, ${p}_{aug}$ and ${p}_{oct}$ reflect the probability to have, at this time, a good estimator of cropland area $\hat{c}$, crop area $\hat{a}$ and crop yield $\hat{y}$, using EO data. These are given for clarity and ease of interpretation. The important point is the temporal sequence in which each estimator is available.

\subsection{Accuracy Requirements}

During the growing season, a  newly available estimator should be used only if it increases the accuracy of production prediction. Therefore, simple rules can be set to define the accuracy requirements of the early estimators $\hat{c}_{t}$,  $\hat{a}_{it}$, $\hat{y}_{it}$ :



\begin{multline}
\mathrm{CV(RMSE_{\textit{may, i}})} \geq \mathrm{CV(RMSE_{\textit{jul, i}})} \geq \mathrm{CV(RMSE_{\textit{aug, i}})} \geq  \\\ \mathrm{CV(RMSE_{\textit{oct, i}})} \geq  \mathrm{CV(RMSE_{\textit{sep, i}})} \geq 0
\label{Eq:RMSEerrorTempSeq}
\end{multline}

From this temporal sequence, it may be inferred that the CV(RMSE) of pre-harvest production estimation (i.e. before September in Senegal) will never be lower than $\mathrm{CV(RMSE_{\textit{sep}})}$ and should not be higher than $\mathrm{CV(RMSE_{\textit{may}})}$. It means that the main factor limiting the accuracy of pre-harvest production estimates is the accuracy of $\hat{y}_{it|T}$ which itself only depends on the inter-annual variability of the yield and its predictability by a trend.   \\

$\mathrm{CV(RMSE_{\textit{may, i}})}$ and $\mathrm{CV(RMSE_{\textit{sep, i}})}$ are independent from the accuracy of the estimators $\hat{c}$, $\hat{a}$ or $\hat{y}$ and only depend on historical data. On the other hand, $\mathrm{CV(RMSE_{\textit{jul, i}})}$ and $\mathrm{CV(RMSE_{\textit{aug, i}})}$ depend on the accuracy of $\hat{c}$ and $\hat{a}$ respectively. It means that when both estimators are available (from August), the best estimator between $\hat{p}_{jul}$ or $\hat{p}_{aug}$ depends on the combined accuracy of $\hat{c}$ and $\hat{a}$. The implications of this specific case on the accuracy requirements are explored in the results.

\section{Results}

\subsection{Trend Analysis}

As expected, the trend (exponential) is highly significant for rice and explains most of the variance of its production, area, and yield (Table \ref{tab:trend}). It is a long-standing objective of the government to become self-sufficient in rice production to decrease the importation costs (see Data section). Several supporting programs and projects have been dedicated to this crop over the years leading to an increase in area and yield and, therefore, production. 
While the impact is less important, a significant exponential trend is also observed for cassava and maize. The trend in cassava production seems to be more explained by a yield increase while for maize, the trend is driven by a rise in cultivated areas. These two crops have also been supported by public policies to diversify the subsistence agriculture offering.

\begin{table}[b!]
\centering
\begin{tabular}{lcccccc}
   &  \multicolumn{2}{c}{Production}  & \multicolumn{2}{c}{Area} & \multicolumn{2}{c}{Yield}\\ 
          \cmidrule(lr{0.5mm}){2-3}       \cmidrule(lr{0.5mm}){4-5}       \cmidrule(lr{0.5mm}){6-7}
 & $\beta$ &  $R^{2}$ & $\beta$  &  $R^{2}$ & $\beta$ &  $R^{2}$ \\ 
\cmidrule(lr{0.5mm}){2-2}\cmidrule(lr{0.5mm}){3-3}\cmidrule(lr{0.5mm}){4-4}\cmidrule(lr{0.5mm}){5-5}\cmidrule(lr{0.5mm}){6-6}\cmidrule(lr{0.5mm}){7-7}\\
Groundnuts & 13.64  & 0.09 & {16.2} &  {0.23} & -0.86 &  0.00 \\ 
  Millet & \underline{0.02} & 0.12 & -1.46 &  0.00 & {10.71} & {0.26} \\ 
  Rice & \underline{\textbf{0.08}} &  \textbf{0.80} & \underline{\textbf{0.05}} & \textbf{0.61} & \underline{\textbf{0.03}} &  \textbf{0.69} \\ 
  Cassava & \underline{\textbf{0.09}} &  \textbf{0.47} & \underline{0.04} &  {0.23} & \underline{\textbf{0.05}} & \textbf{0.55} \\ 
  Maize & \underline{\textbf{0.07}} &  \textbf{0.39} & \underline{\textbf{0.05}} &  \textbf{0.58} & \underline{0.02} &  0.13 \\ 
  Sorghum & 0.91 & 0.02 & \underline{-0.01} & 0.02 & 7.35 & 0.12 \\ 
  Cotton & -0.43 &  0.04 & {-0.76}  & {0.22} & 5.15 & 0.02 \\ 
\end{tabular}
    \caption{Results of trend analysis using OLS regressions (linear or \underline{exponential}) performed over historical production, area and yield (1997-2016). Slope ($\beta$) with a $p$-value $<0.01$, marked in bold, were considered significant.}
    \label{tab:trend}
\end{table}

\subsection{Lowest Error of Early Estimation of Production}

\begin{figure}[t!]
    \centering        
        \includegraphics[width=0.6\linewidth]{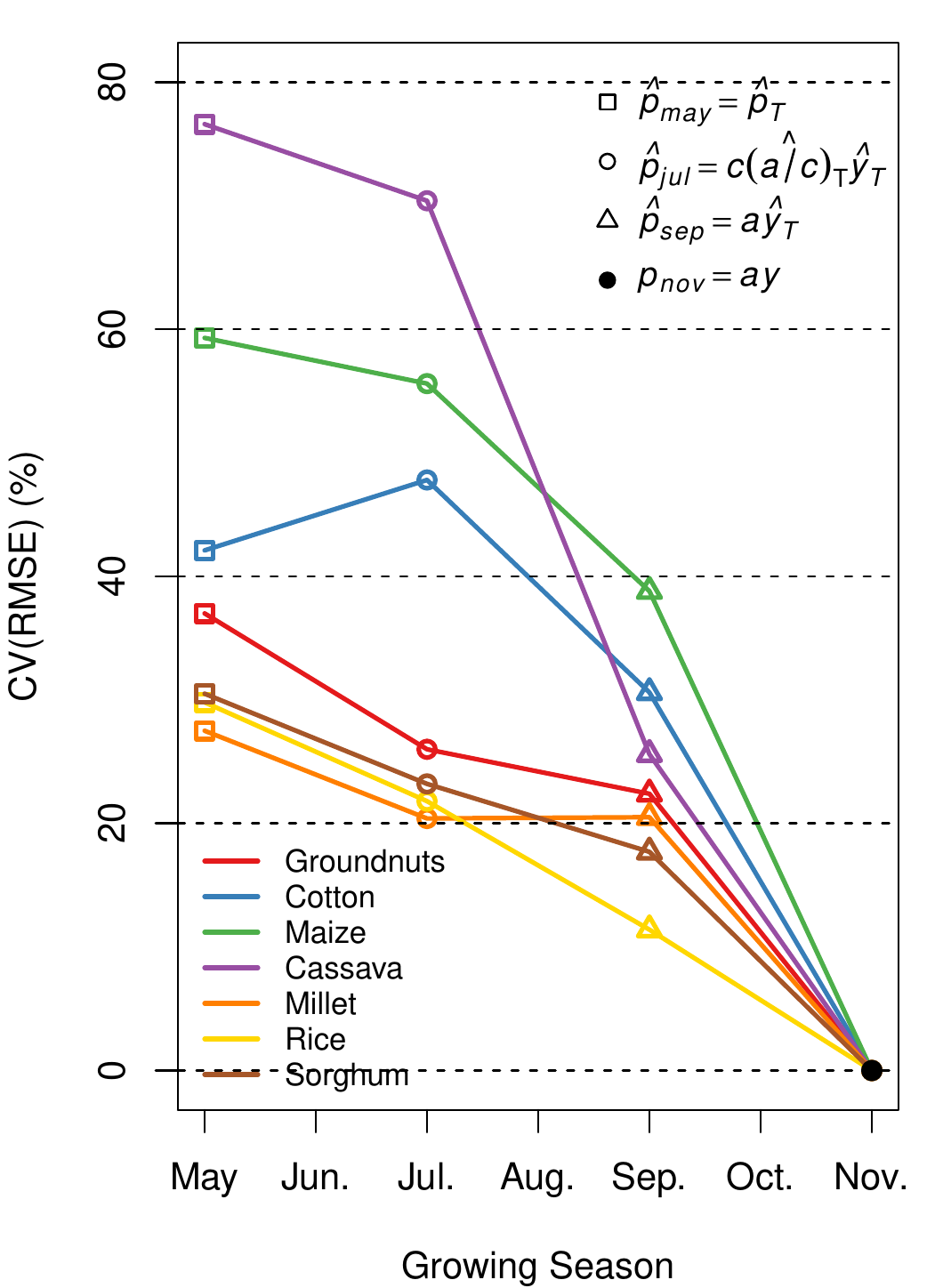}   
    \caption{Lowest error of production estimation, expressed in CV(RMSE), achievable for each crop according to the increasing data availability along the season (considering perfect estimator of cropland area, $\hat{c}=c$). }
    \label{fig:calendarV2}
\end{figure}

Figure \ref{fig:calendarV2} shows the lowest error of production estimation (expressed in CV(RMSE)) achievable for each crop according to the increasing data availability along the season (considering a perfect estimator of cropland area, $\hat{c}=c$)). \\

At the beginning of the season in May, production estimates are given by the historical average (and the trend, if any) of past production data ($\hat{p}_{may}$). The $\mathrm{CV(RMSE_{\textit{may}})}$ at this time depends on the inter-annual variability of production and its predictability by a trend. Despite their significant trend, cassava and maize have the highest $\mathrm{CV(RMSE_{\textit{may}})}$ (77\% and 59\% respectively) due to the high inter-annual variability of their production (CV of 86.2\% and 60.2\% respectively, see Table \ref{tab:statDesc}). All others variables have a $\mathrm{CV(RMSE_{\textit{may}})}$ in line with their CV except for rice due to its significant trend ($\mathrm{CV(RMSE)} = 30\%$ and $\mathrm{CV} = 60\%$). The crop with the lowest $\mathrm{CV(RMSE_{\textit{may}})}$ is millet ($27\%$) because of its relatively lower inter-annual variability. \\

Having a perfect estimator of cropland decreases the error of production estimation for all crops apart from cotton for which an estimator of cropland, even perfect, is virtually useless. It means that the cropland area alone is already a valuable information to improve early production estimation. Groundnuts, sorghum, rice and millet have all a $\mathrm{CV(RMSE_{\textit{aug}})}$ in the  20-26\% range. However, for highly variable crops (cassava and maize), the error of estimation is still large. Note that in Figure \ref{fig:calendarV2}, we arbitrary set the date at which a perfect cropland would be known (in July). However, the actual cropland is only known in September at the same time than crop area  (when it becomes useless). It is, therefore, just an indication of what error is expected before September, if a perfect estimation of cropland area could be obtained. Practically, as a perfect early estimator of cropland does not exist, the $\mathrm{CV(RMSE_{\textit{aug}})}$ achievable will be in-between the one of $\hat{p}_{may}$ and $\hat{p}_{jul}$ depending on the accuracy of the cropland estimation (see section 1.5.4). \\

In September, the official statistics of crop area are known (Figure \ref{fig:SchemeCalendar} -- Official Statistics). The $\mathrm{CV(RMSE_{\textit{may}})}$ and  $\mathrm{CV(RMSE_{\textit{sep}})}$ define the range of error of pre-harvest production estimation. The error should not be larger than $\mathrm{CV(RMSE_{may})}$ and cannot be smaller than $\mathrm{CV(RMSE_{sep})}$ because the yield cannot theoretically be computed before the peak of the growing season in September (Figure \ref{fig:SchemeCalendar} -- Early Estimators). All crops have a $\mathrm{CV(RMSE_{\textit{sep}})}$ below 26\% apart from cotton and maize (representing less than 12\% of the total agricultural production in Senegal, see Figure \ref{fig:Pie}). This result is completely independent from
the accuracy of early estimators of cropland area or crop area.  \\

Finally, there is a period between September and November when area is known and yield can be estimated with EO data. A good estimator of crop yield can substantially decrease the error of production estimates. During this period, the best prediction of production can be achieved. 

\subsection{Spatial Variability}

\begin{figure}
    \centering        
        \includegraphics[width=0.95\linewidth]{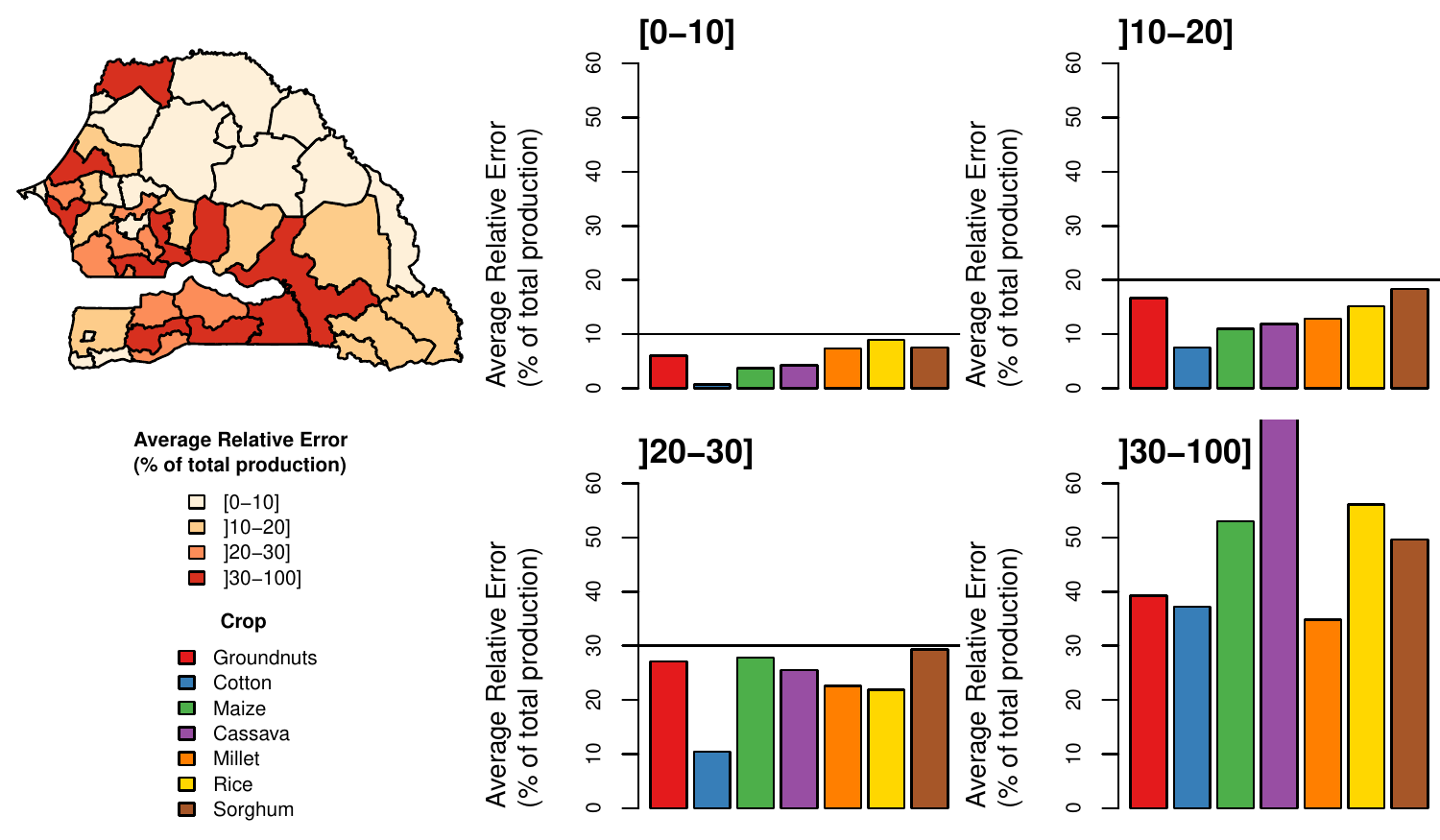}         
    \caption{Stratification of Senegal based on average $\hat{p}_{may}$ error by department (expressed in percentage of national production) in each department (data from 2010 to 2016). Strata are shown on the map while the average production error per crop and for each stratum is depicted by the bar plots. The number of departments per stratum is 15 in [0-10], 10 in ]10-20], 8 in ]20-30], and 10 in ]30-100].}
    \label{fig:Cost}
\end{figure}

Figure \ref{fig:Cost} shows a stratification of Senegal based on $\hat{p}_{may}$ error in each department (using 2010-2016 data). It illustrates the spatial distribution of the production variability within the country. If the production was known in 65\% of all departments and estimated using $\hat{p}_{may}$ in the remaining 35\% (stratum [0-10]), the expected error of total production would be, in average, less than 10\% for each crop. For a maximal average error of 20\%, the number of departments that you could estimate with $\hat{p}_{may}$ jumps to 58\%. It means that around 40\% of the departments explained most of the national inter-annual variability of production for all crops. These departments are also the most productive (pearson correlation of 0.79). (Figure \ref{fig:MapProd}). This stratification of the country could be used to reallocate the sampling effort to departments with high variability.

\subsection{Accuracy Requirements of Cropland Area, Crop Area and Crop Yield}

Table \ref{tab:accReq} presents the accuracy requirements (expressed in maximum relative error) of cropland area, crop area and crop yield computed using the rules defined in Eq. \ref{Eq:RMSEerrorTempSeq}. As already mentioned in the previous sections, knowing the cropland area is useless for cotton as $\hat{p}_{jul}$ gives higher errors of production estimation than $\hat{p}_{may}$. On the other hand, the accuracy requirement for the cropland is low to estimate cassava production but, as shown in Figure \ref{fig:calendarV2}, the error of $\hat{p}_{jul}$ is still high. For the other crops, the error should be lower than 17-24\%. A cropland classification based on EO data should have at least an overall accuracy of 83\% to provide estimation of production more accurate than the historical average and trend ($\hat{p}_{may}$) for all crops. Below 76\% of accuracy, such an estimator would be useless as it would not improve the production prediction of any crops compared to $\hat{p}_{may}$ (excluding cassava). \\

For crop area, the maximum error allowed is a bit higher ranging from 17 to 50\%. Excluding cassava, the minimum accuracy of a crop area classification is 65\% (and 73\% by excluding maize and cassava). And the minimum accuracy to improve each crop is again 83\%. Cassava has lower requirements because the $\mathrm{CV(RMSE_{may})}$ was very high for this crop. It is therefore easy to improve its production prediction even with a poor estimator of crop area. \\

Finally for crop yield, the requirements ranges from an error of 10 to 34\%. The lower the $\mathrm{CV(RMSE_{sep})}$, the higher is the requirement. A very accurate estimator of crop yield is needed to improve production estimation of rice and sorghum after September (maximum error of 10 and 13\% respectively) while estimates for maize and cotton can be improved easier (maximum error of 34 and 28\% respectively).

\begin{table}[b!]
\centering
\begin{tabular}{lccccccc}

  &  Groundnuts & Millet & Rice & Cassava & Maize & Sorghum & Cotton \\
\midrule
  cropland area, $\hat{c}$   &  24\% & 17\% & 20\%  & > 50\% & 17\% & 17\%  & -  \\
    crop area, $\hat{a}$   & 27\% & 17\% & 23\%  & 50\% & 35\% &  24\% & 27\%  \\
  crop yield, $\hat{y}$   & 21\% & 20\% &  10\% & 20\% & 34\% &  13\% &  28\% \\
\bottomrule

\end{tabular}
    \caption{Accuracy requirements of cropland area, crop area and crop yield for each crop, expressed in maximum error of estimation (\%). The requirements takes the worst case scenario between underestimation and overestimation.}
    \label{tab:accReq}
\end{table}

\subsection{Combined Requirements for Cropland Area and Crop Area Estimators}

\begin{figure}
    \centering        
        \includegraphics[width=1\linewidth]{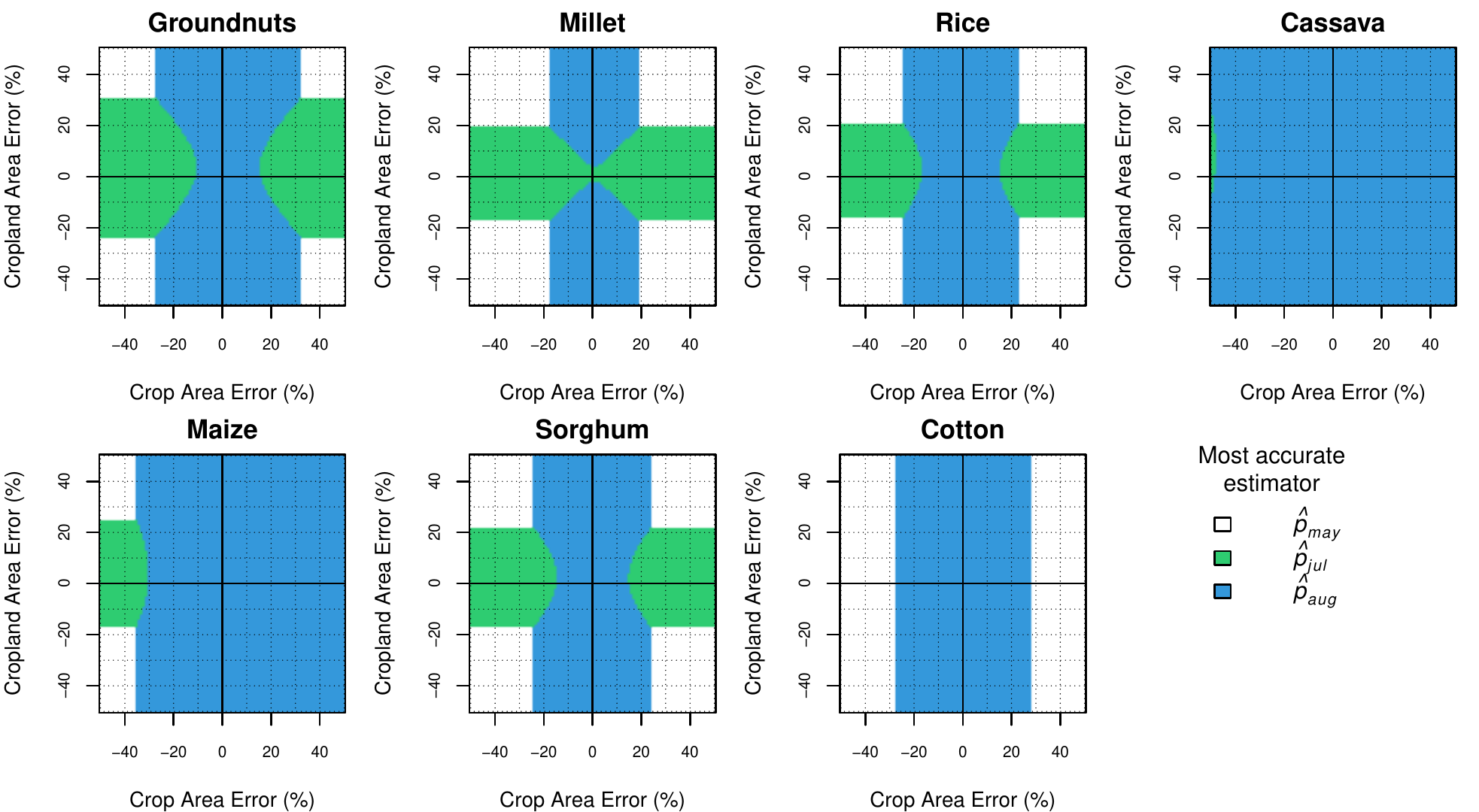}         
    \caption{
    Most accurate estimators of crop production before September between $\hat{p}_{may}$, $\hat{p}_{jul}$ and $\hat{p}_{aug}$ according to the error of estimators of cropland $\hat{c}$ and crop area $\hat{a}$.}
    \label{fig:priorArea}
\end{figure}

Figure \ref{fig:priorArea} shows which estimator gives the lowest CV(RMSE) between $\hat{p}_{may}$, $\hat{p}_{jul}$ and $\hat{p}_{aug}$ according to the accuracy of the estimators of cropland area $\hat{c}$ and crop area $\hat{a}$. In other words, this figure indicates which estimator, between $\hat{p}_{jul}$ and $\hat{p}_{aug}$, should primarily be used to estimate production before September. This case can be observed from August when both $\hat{p}_{jul}$ and $\hat{p}_{aug}$ are available. It could also potentially be before August if an accurate enough estimator of crop area is available at that time.  \\

The boundaries of the white areas indicate the accuracy requirements for $\hat{p}_{jul}$ and $\hat{p}_{aug}$ reported in Table \ref{tab:accReq}. The whiter a plot, the harder it is to get better estimation of production than $\hat{p}_{may}$. It reflects the low variability of production and/or its predictability by a trend.\\

Again, the figure shows that cropland is useless for cotton and virtually useless for cassava as even poor estimates of crop area (error > 50\%) already give better results than $\hat{p}_{jul}$ no matter the accuracy of $\hat{c}$. For the same error, it is always better to use an estimator of crop area rather than of cropland to predict the production, i.e., $\mathrm{CV(RMSE_{jul})}$ is always lower than $\mathrm{CV(RMSE_{aug})}$ for the same accuracy of $\hat{c}$ and $\hat{a}$. Millet is an exception as for $\hat{c}$ and $\hat{a}$ with identical accuracy, the two estimators $\hat{p}_{jul}$  and $\hat{p}_{aug}$ provide the same results as shown by the 1:1 diagonal boundary between blue and green areas. For groundnuts (and millet), if a very accurate estimator of $\hat{c}$ is available, it might be better to estimate the production with $\hat{p}_{jul}$ rather than $\hat{p}_{aug}$. The effect is less strong for rice and sorghum. \\

Production is computed from the multiplication of two separated estimators ($\hat{a}$ and $\hat{y}$). The errors of $\hat{a}$ and $\hat{y}$ can either compensate for each other (underestimation times overestimation), or multiply each other (underestimation times underestimation or overestimation times overestimation). This fact means that if, for instance, $\hat{y}$ overestimates the yield then the error of $\hat{p}$  will be lower if  $\hat{a}$ underestimates the area. This inevitably leads to asymmetric accuracy requirements depending on the potential compensation of errors. As shown on Figure \ref{fig:priorArea}, this asymmetry has not a strong effect except for maize ($\hat{p}_{aug}$). This is explained by the yields in 2003 to 2005 that are much higher than the average and are therefore underestimated by $\hat{y_{t|T}}$. For this crop, an overestimation of $\hat{a}$ gives better production estimation than an underestimation for the same absolute error.


\subsection{Combined Requirements for Crop Yield and Crop Area Estimators}

\begin{figure}
    \centering        
        \includegraphics[width=1\linewidth]{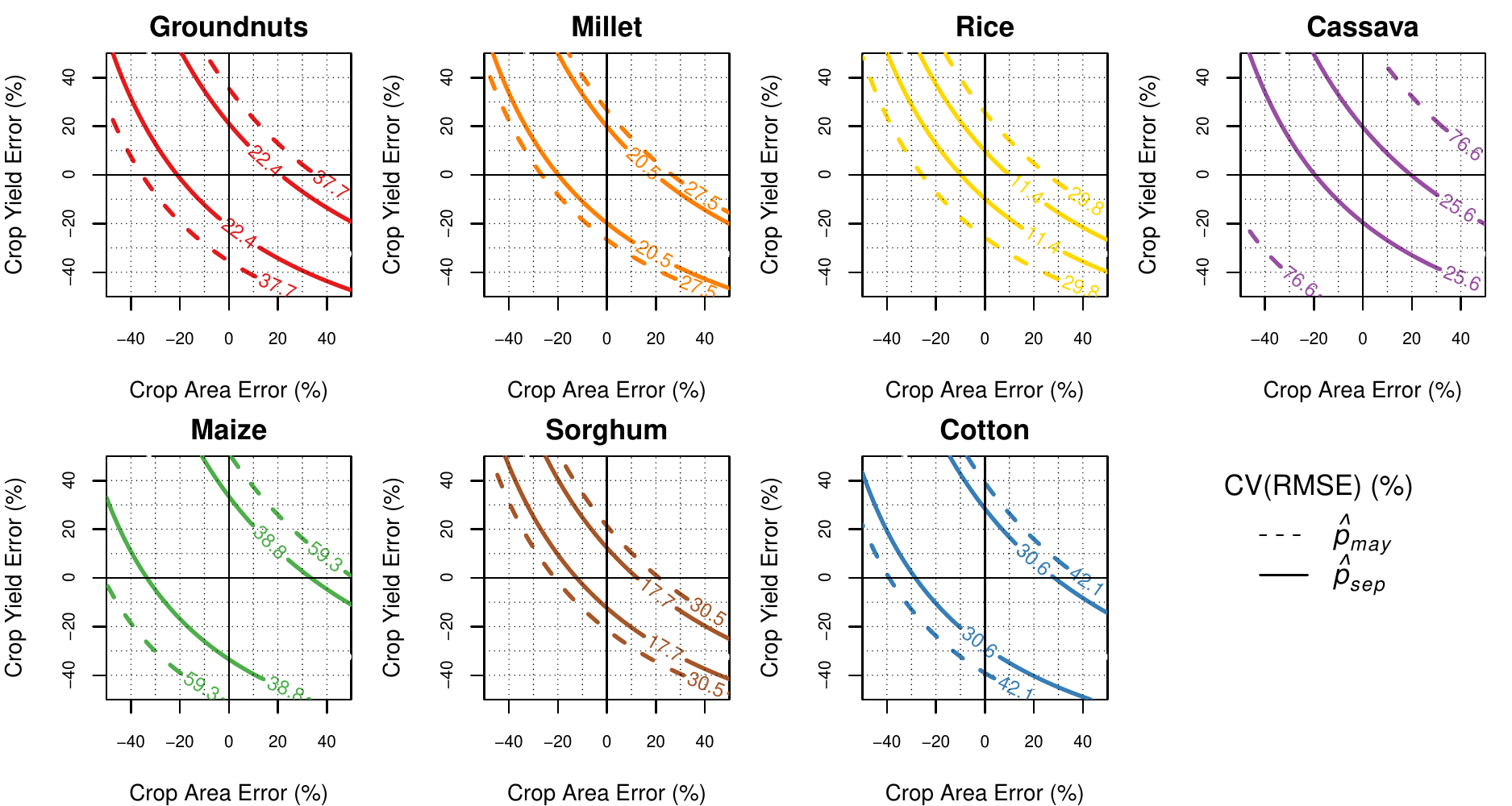}         
    \caption{
    Production accuracy, expressed in CV(RMSE) (\%), achievable for a combination of yield and area error for each crop. The isolines show the combination of yield and area error giving the same CV(RMSE) than $\hat{p}_{may}$ (dotted lines) and $\hat{p}_{june}$ (full lines). To be useful the combination of errors of the estimators should give a CV(RMSE) that falls within the limits of the dotted lines before September and within the limits of the full lines after September with crop area error = 0\% at this date.}
    \label{fig:AreaYieldError}
\end{figure}

It has already been mentioned that it was very unlikely to get accurate estimations of yield before the end of the growing season (based on EO data). However, an exact estimator is not necessarily needed. Indeed, any estimator that would provide more precise estimation than $\hat{y}_{it|T}$ can help achieve a more accurate estimation of production than $\hat{p}_{sep}$. Figure \ref{fig:AreaYieldError} shows the accuracy requirements of crop yield to get lower production error than $\hat{p}_{may}$ (dotted lines) and $\hat{p}_{sep}$ (full lines) depending on the accuracy of crop area, and vice versa. For the same accuracy, two underestimated components always give lower errors of production than two overestimated (due to the multiplication of errors). On the other hand, the compensation of errors (when one component is underestimated and the other overestimated) can highly decrease the accuracy requirements of $\hat{y}$ and $\hat{a}$. For early warning, only the worst case scenario, when the two components were overestimated (top right quadrants in Figure \ref{fig:AreaYieldError}), should be considered. When the area is known in September, the accuracy requirements of crop yield are given by the intersection of $\hat{p}_{sep}$ isoline with the vertical line corresponding to an error of 0\% for crop area (the values are reported in Table \ref{tab:accReq}). 

\section{Discussion}

Due to supportive programs, significant trends were observed in rice, cassava and maize production (Table \ref{tab:trend}). The trend was substantial for rice by explaining 80\% of the production variance. While trend can explain most of the variability of some crops, they generally result from specific policies that can abruptly change from year to year (e.g. due to specific policy). Furthermore, the trend cannot increase indefinitely as biophysical factors limit the yield and the extent of the cultivated areas. Consequently, accounting for the trend for early warning should always be made in light of the current political context and the environmental constraints. More extended time series would also strengthen the analysis. In particular, a long-term trend coming from progressive technical improvement (new equipment, new crop varieties, etc.) would have been easier to model.\\

From the results, accuracy requirements for early estimators of cropland area, crop area and crop yield can be defined by two key values: the maximum error to improve the estimation of production of all crops (17\%, 17\%, 10\% respectively) and the maximum error to improve the estimation of production of at least one crop (24\%, 27\%, 34\% respectively). We exclude cassava for cropland area and cassava and maize for crop area because the error of production, CV(RMSE), remain large at lower requirements. Not that for cassava, early estimates of production might not really make sense because the seeding and harvesting of this crop do not necessarily occurred the same year. Therefore, the number reported for this crop should be interpreted with caution. Estimators of cropland area can be used to improve the accuracy of early production estimates of groundnuts, millet and rice, the three main crops in Senegal stressing the value of cropland mapping for food security. To be able to use these requirements operationally, the expected accuracy of $\hat{c}$, $\hat{a}$ and $\hat{y}$ should be known before the beginning of the season. This requires a preliminary benchmark of each estimator using historical data.  \\ 


Because the actual area is known at the beginning of the harvest (Figure \ref{fig:SchemeCalendar}), the inter-annual variability of crop yield is the main factor limiting the accuracy of pre-harvest production estimates. The underlying assumption is that the crop yield is not predictable before the harvest. In the Sudano-Sahelian zone, production variability is generally associated with rainfall variability because agriculture is mostly water-limited \citep{d2015senegal}. The yield of rain-fed crops has been shown many times to be correlated with accumulated rainfall \citep{dennett1981rainfall, sultan2013assessing}. In Figure \ref{fig:rainfall_analysis} in Appendix, we show the correlation between cumulative rainfall over the rainy season and the crop yield. Groundnuts, sorghum, and millet were poorly correlated before September while maize, cotton and cassava showed higher correlation for July.  This early correlation for maize, cotton and cassava could be used to improve pre-harvest production estimation. \\


Early estimations of cropland and crop area can be obtained from the classification of Earth Observation images. The downside of this approach is that the best methods of classification tend to rely on supervised algorithm calibrated with training data. To be efficient, these data should be collected using a sound statistical sampling, directly on the ground (or by photo-interpretation, but this approach requires very high resolution due to the fragmentation of the cropland landscape in Africa). In the context of underfunded NSO, there is little argument to support the organization of two field campaigns (for calibration and official statistics) collecting the exact same information (even tough a smaller sample is needed to calibrate classification algorithms). On the other hand, some classification methods do not require unbiased calibration sample (e.g. maximum likelihood). Furthermore, several methods have been developed to automatically derive a calibration sample using past data, such as the crop map of the previous year, as reference 
\citep{matton2015automated, waldner2017national}. Note that for crop yield, simulated data from agro-meteorological models have also been used as calibration set \citep{burke2017satellite,lobell2015scalable}. Therefore, emphasis should be put on the development of such methods rather than on the one relying on ground data. On the other hand, EO data allow to produce exhaustive map of crop areas and yield. This information has an added value compared to simple agricultural statistics for early warning systems because it allows to precisely locate the areas with potential crop failures. Furthermore, if major events occur such as a drought or locust invasion, timeliness can be more critical than very accurate estimates. In these cases, photo-interpretation or classification of Earth Observation images can provide precious information for an early diagnosis of the situation \citep{renier2015dynamic, rhee2010monitoring}.\\

In this study, crop statistics provided by national offices were considered as the ground truth. However, it only provides another estimation of production impacted by a sampling bias and variance \citep{stehman2005comparing}. In Senegal, the quality of the statistical capacities are rather good compared to the African average (see Introduction). The two-stage stratified sampling limits the sampling variance and if a bias exists, it is likely to be similar from year to year since the sampling method is always the same. \\



We used the CV(RMSE) as the error measurement of production estimators. However, the distribution of errors can be very different for the same value of CV(RMSE) (see Figure \ref{fig:RMSEdif} in Appendix). The interpretation of such an indicator is, therefore, not necessarily straightforward  and prevents the appropriate management of early warning information by decision makers. In a paper on early warning systems for food security in West Africa, \citet{genesio2011early} concluded that a better tailoring of information is needed as the interpretation of forecasts indicators is not well understood by the decision makers. In reporting statistical indicators, it is essential to consider the intended audience and their understanding of their statistical meaning as a misinterpretation of information can subsequently impact on decision making \citep{wallach2015uncertainty, budescu2009improving, morton2011future}. \citet{genesio2011early} also stressed the fact that politically, it is easier to deal with the effect of a disaster than to take action on the basis of uncertain information.\\

Finally, food availability is only one component of food security. For early warning system, monitoring access to food is capital as it is often regarded as more important than food availability \citep{sen1981poverty}. Access is conditioned by several factors including staple prices and household income for which EO data can also provide interesting insight \citep{jacques2018social, pokhriyal2017combining}.

\section{Conclusions}

This work proposed a methodological framework to define accuracy requirement for early warning estimators of crop production according to the inter-annual variability of historical data. The analysis was applied to the seven main crops in Senegal, a country with highly variable agricultural production, using time series of twenty years of production data.\\

We showed that the inter-annual variability of crop yield was the main factor limiting the accuracy of pre-harvest production estimates because the actual crop area is always known earlier than the crop yield in Senegal. The lower the yearly variability of the yield, or the easier this variability is predictable (e.g. by a trend), the more accurate could be the pre-harvest estimation of production for a specific crop. Interestingly, estimators of cropland area were useful to improve production predictions of the main crops in Senegal stressing the value of cropland mapping for food security. \\

That being said, apart from rice (mainly predict by the trend), the lowest CV(RMSE) achievable before the harvest (considering a perfect estimator of crop area) were rather high ranging from 18\% to 40\%. Therefore, quantitatively estimate production for early warning using E0 data might be a futile attempt. On the other hand, EO data are still highly relevant as a source of qualitative information for early warning by, for instance, identifying strong anomalies in vegetation index or rainfall pattern. They also provide precise and exhaustive geospatial information at high frequency which is not given by simple crop statistics. Furthermore, they constitute sometimes the only source of information available in remote or unsecured areas. In another vein, alternative data sources such as citizen science/crowdsourcing might be a promising bottom up approach to get early indicators of food security directly from the ground \citep{minet2017crowdsourcing}. \\



This analysis was based on the premise that estimates from official statistics were robust and could be considered as the ground truth data. This assumption might be strong as sampling, and non-sampling errors can have a great impact on the accuracy of such statistics. Therefore, the inter-annual variability of production and consequently, the accuracy requirements, could be wrongly estimated. A depth assessment of the quality of the estimation provided by official statistics should, therefore, be carried out first before firmly relying on the outputs of such analysis. More robust results are also expected from more extended time series and analyses at the local scale.\\

The framework applied in this study is the first step to further research on the inter-annual variability of production in the context of early warning. It stresses the importance of better tailoring early estimators of production in line with the variability of historical data and the calendar of official statistics data collection. 

\vspace{6pt} 



\authorcontributions{For research articles with several authors, a short paragraph specifying their individual contributions must be provided. The following statements should be used “conceptualization, X.X. and Y.Y.; methodology, X.X.; software, X.X.; validation, X.X., Y.Y. and Z.Z.; formal analysis, X.X.; investigation, X.X.; resources, X.X.; data curation, X.X.; writing—original draft preparation, X.X.; writing—review and editing, X.X.; visualization, X.X.; supervision, X.X.; project administration, X.X.; funding acquisition, Y.Y.”, please turn to the  \href{http://img.mdpi.org/data/contributor-role-instruction.pdf}{CRediT taxonomy} for the term explanation. Authorship must be limited to those who have contributed substantially to the work reported.}

\funding{This work was funded by the Belgian National Fund for Scientific Research through Fonds pour la Formation à la Recherche dans l’Industrie et dans l’Agriculture (Grant 5211815F).}

\acknowledgments{In this section you can acknowledge any support given which is not covered by the author contribution or funding sections. This may include administrative and technical support, or donations in kind (e.g., materials used for experiments).}

\conflictsofinterest{The authors declare no conflict of interest.} 

\abbreviations{The following abbreviations are used in this manuscript:\\

\noindent 
\begin{tabular}{@{}ll}
MDPI & Multidisciplinary Digital Publishing Institute\\
DOAJ & Directory of open access journals\\
TLA & Three letter acronym\\
LD & linear dichroism
\end{tabular}}



\newpage

\reftitle{References}




\appendixtitles{no} 
\appendixsections{multiple} 
\newpage
\appendix
\section{Correlation between Rainfall and Crop Yield}
 


\begin{figure}[h]
    \centering        
        \includegraphics[width=1\linewidth]{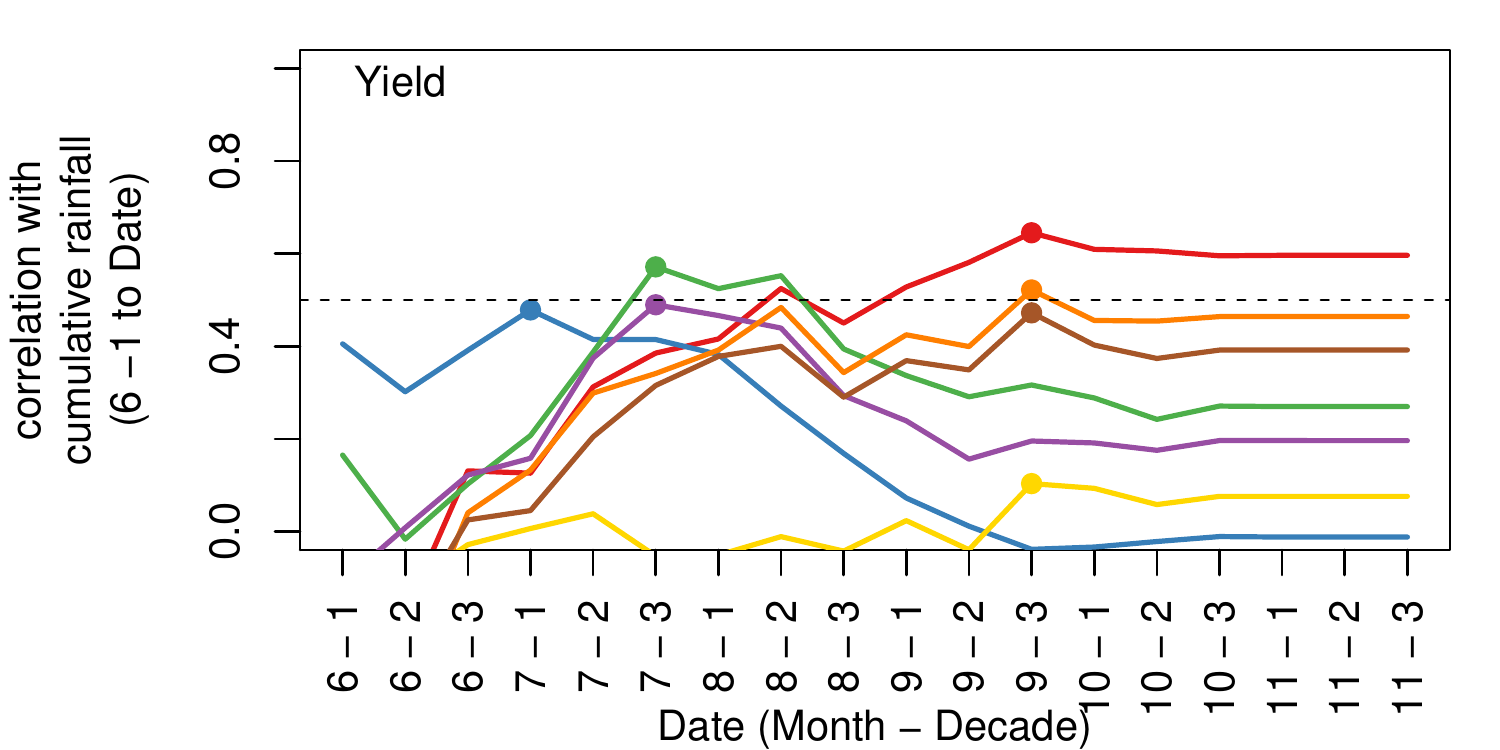}    
    \caption{Correlation between crop yield and average accumulated rainfall over the country (from the first decade of June to each decade of the rainy season) for 20 years (1997-2016). All correlation higher than 0.5 (dotted line) were statistically significant ($p$-value<0.01). For color code, refer to Figure \ref{fig:Pie}.}
    \label{fig:rainfall_analysis}
\end{figure}

The two interesting rainfall features to predict crop yield variability at the country level are the cumulative annual rainfall variability and temporal characteristics (frequency and intensity) \citep{berg2010dominant}. Here, average cumulative rainfall over the country was used. Figure \ref{fig:rainfall_analysis} shows the correlation between cumulative rainfall over the rainy season and the crop yield. The most correlated period depends on each crop. \\

Most of the rice is irrigated in Senegal River Valley, and its dependence on rainfall is therefore limited which is illustrated by the very low correlation observed for this crop. Cotton is mainly cultivated in the South of the country (Figure \ref{fig:MapProd}) where the rainy season starts early compared to the rest of the country. This could explain the higher correlation observed for this crop early in the season (July). \\ 

Groundnuts, sorghum, and millet were better estimated later in the season (September) while maize and cassava showed higher correlation for the end of July.  The correlations observed for the yield were in line with previous results in Senegal \citep{krishnamurthy2013climate}. For the area, an unexpected high correlation was found for millet ($\sim$0.77). \\

Rainfall data were downloaded from the seasonal explorer developed by the Vulnerability Assessment and Mapping of the World Food Programme. The primary data source is the CHIRPS gridded rainfall dataset produced by the Climate Hazards Group at the University of California, Santa Barbara. 
CHIRPS is a 35+ year quasi-global rainfall dataset incorporating satellite imagery with in-situ station data to create gridded rainfall time series. Full details on the underlying methodology can be found in \citep{funk2015climate}. CHIRPS data were shown to have low systematic errors (bias) and low mean absolute errors \citep{peterson2013climate}. CHIRPS is free to use and accessible online (\url{http://chg.geog.ucsb.edu/data/chirps/}). 


\subsection{Distribution of errors for identical CV(RMSE)}

\begin{figure}[h]
    \centering
		\includegraphics[width=1\linewidth]{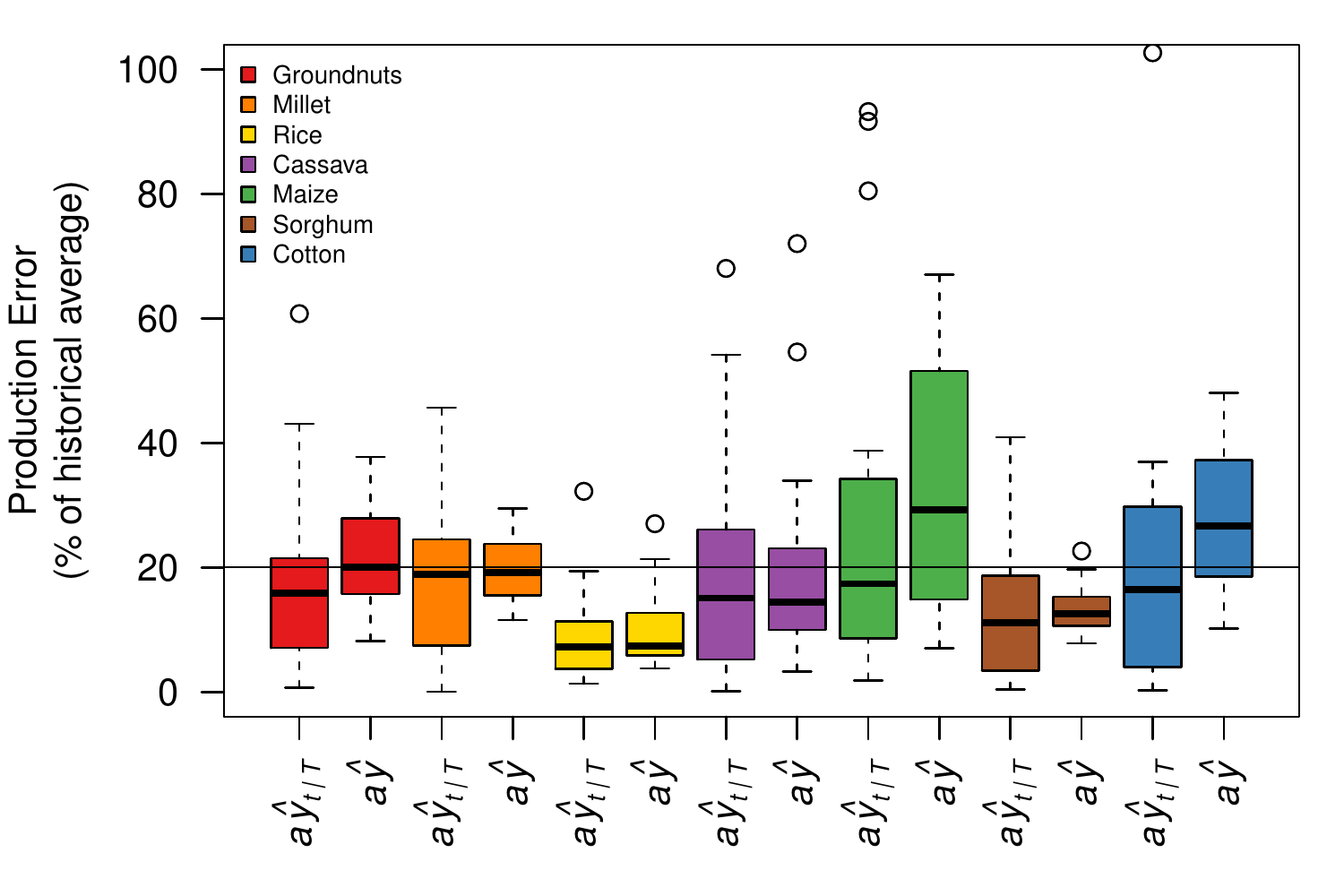}
    \caption{Box-plots of the distribution of production errors (\% of historical average) for $\hat{p}_{aug} = a\hat{y}_{t|T}$ and $\hat{p}_{aug} = a\hat{y}$ where the accuracy of $\hat{y}$ corresponds to the accuracy requirement of crop yield defined in Table \ref{tab:accReq}. Both distribution give the exact same CV(RMSE). However, $a\hat{y}$ has a higher median value but with lower number of outliers.}
    \label{fig:RMSEdif}
\end{figure}

\unskip


\end{document}